\documentclass[aps,twocolumn,showpacs,preprintnumbers]{revtex4}
\usepackage{dcolumn}
\usepackage{bm}
\usepackage{epsfig}

\begin{document}
\title{Chiral Condensates in Quark and nuclear Matter
\footnote{Respective email addresses of the authors:\\
           G.X.Peng,   gxpeng@ihep.ac.cn;
           U.Lombardo, lombardo@lns.infn.it;
           M.Loewe,    mloewe@fis.puc.cl;
           H.C.Chiang, chiang@ihep.ac.cn;
           P.Z.Ning,   ningpz@nankai.edu.cn}
      }
\author{G.\ X.\ Peng,$^{1,2,3}$ U.\ Lombardo,$^{4}$ M.\ Loewe,$^5$
        H.\ C.\ Chiang,$^2$ and P.\ Z.\ Ning$^3$
             }
\affiliation{
 $^1$China Center of Advanced Science and Technology (World Lab.),
               Beijing 100080, China\\
 $^2$Institute of High Energy Physics, Chinese Academy of Sciences,
               Beijing 100039, China\\
 $^3$Department of Physics, Nankai University, Tianjin 300071, China\\
 $^4$Dipartimento di Fisica and INFN-LNS,
               Via S.\ Sofia 44, 95123 Catania, Italy\\
 $^5$Facultad de Fisica, Pontificia Universidad Cat\'{o}lica de Chile,
               Santiago 22, Chile
                    }
\begin{abstract}
We present a novel treatment for calculating the in-medium quark condensates.
The advantage of this approach is that one does not need to make further
assumptions on the derivatives of model parameters with respect to the quark
current mass. The normally accepted model-independent result in nuclear
matter is naturally reproduced. The change of the quark condensate induced by
interactions depends on the incompressibility of nuclear matter. When it is
greater than 260 MeV, the density at which the condensate vanishes is higher
than that from the linear extrapolation. For the chiral condensate in quark
matter, a similar model-independent linear behavior is found at lower
densities, which means that the decreasing speed of the condensate in quark
matter is merely half of that in nuclear matter if the pion-nucleon sigma
commutator is six times the average current mass of $u$ and $d$ quarks.
The modification due to QCD-like interactions is found to slow the
decreasing speed of the condensate, compared with the linear extrapolation.
\end{abstract}

\pacs{21.65.+f, 24.85.+p, 12.38.-t, 11.30.Rd}

\maketitle

\section{Introduction}

 The spontaneously breaking of chiral symmetry in Quantum
Chromodynamics (QCD) has been an interesting topic for nuclear
physics~\cite{bro01bro95}.
It is expected that this broken symmetry would be restored in a medium,
and several signals have recently been suggested for its manifestation
in nuclear matter, such as the spectral enhancement near the $2M_{\pi}$
($M_{\pi}$ is the pion mass) threshold~\cite{Hatsuda99}, the systematic
appearance of the parity doublets in the missing-states
region~\cite{Glozman01},
 and the vector manifestation~\cite{Harada01} etc.
It is also shown that the partial restoration of chiral symmetry has
strong effects on the nuclear saturation~\cite{Rapp99}
and the nuclear equation of state~\cite{Santra00}.

  One of the most relevant quantity to the chiral symmetry restoration
is the scalar quark condensate $\langle\bar{q}q\rangle$. The zero value
of the condensate can be strictly proven to be a necessary condition
for exact chiral symmetry. But the sufficiency is not available presently.
In fact, Birse has shown that the pieces arising from low-momentum virtual
pions should not be associated with chiral
restoration~\cite{Birse}.
However, more recent investigations
are in favor of the hypothesis that the quark condensate is a leading
order parameter~\cite{Cavalcante01,Colangelo01}.

   The quark condensate in vacuum $\langle\bar{q}q\rangle_0$
is indeed a vast quantity ($\sim -10^7$ MeV$^3$). However, this
value will be modified in a medium. The change produced by the
deuteron has recently been investigated~\cite{Ballot00}. There are
also considerable investigations on the modification due to nuclear
and/or quark matter
\cite{Drukarev99,Cohen91,Chanfray93,Lutz,Guohua,Malheiro97%
,Mazy,Mitsu,Lilei,Brock,Fojii,Ligq,Bernard87,Jaminon92}.
For more references, see relevant review
articles~\cite{bro01bro95,Birse94}.

  It has long been speculated that strange quark matter (SQM),
rather than the normal nuclear matter, might be the true QCD ground
state~\cite{Witten84}, or absolutely stable
\cite{Jaffe84,PengPRC56,Cha,Ben,Surk}. The critical density of
SQM depends on the quark mass scaling~\cite{PengPRC59},
while the quark mass scaling is closely linked to the quark condensate
in quark matter~\cite{PengPRC61}.
Therefore, the study of in-medium quark condensates
is meaningful independently of whether it can serve as
an order parameter of chiral restoration.

  Originally, the interacting part of the quark mass was
taken to be inversely proportional to
density~\cite{Cha,Ben,Surk,PengPRC59}.
That parametrization has no derivation, and leads to
inconsistencies in the thermodynamic properties of
SQM~\cite{PengPRC62}.
Later, a straightforward derivation method is provided,
and the scaling is changed to be inversely
proportional to the cubic root of the density~\cite{PengPRC61}.

  One of the most popularly used methods to calculate the
in-medium quark condensate is the
Feynman-Helmann theorem. The main difficulty in this formalism
is that one has to make assumptions on the derivatives of model
parameters with respect to the quark current mass. To bypass this
difficulty, we will apply a similar idea as in the study of
strange quark matter by defining an equivalent mass.
A differential equation which determines the equivalent
mass will be derived in detail, then it is solved within a QCD-like
inter-quark interaction.
The normally accepted linear behavior of the condensate in nuclear
matter is naturally reproduced and extended to quark matter.
However, the decreasing speed of the condensate in quark matter
is merely half of that in nuclear matter if
the pion-nucleon sigma term is 6 times the average current mass of
light quarks. At higher densities, the decreasing speed
is usually slowed, compared with the linear extrapolation,
due to interactions.

   We organize the present paper as follows. In the subsequent
Sec.\ \ref{qcqm}, we describe the equivalent mass method, and apply
it to the investigation of the quark condensate in quark matter.
An equation to determine the equivalent mass is derived,
solved analytically at lower densities,
and solved numerically with a QCD-like inter-quark interaction
at higher densities.  To test the correctness of
the equivalent mass method, we apply it to nuclear matter
in Sec.\ \ref{qcnm}. The model-independent linear behavior is
reproduced and the higher order's contribution is studied in
the mean-field level. Sec.\ \ref{dis} is a short summary.

\section{The chiral condensate in quark matter}
\label{qcqm}

   Applying the Feynman-Hellmann theorem, respectively, to the
vacuum $|0\rangle$\ and the state $|n_{\text{b}}\rangle$ with
baryon number density $n_{\text{b}}$, and then taking the difference,
as has been done for nuclear matter~\cite{Cohen91}, one can obtain the
following expression for the relative chiral condensate in quark matter:
\begin{equation}  \label{qcCohen}
  \frac{\langle\bar{q}q\rangle_{n_{\text{b}}}}{\langle\bar{q}q\rangle_0}
  = 1 - \frac{1}{3n^*}
        \frac{\partial\epsilon}{\partial{m_0}},
\end{equation}
where $\epsilon$
is the energy density of quark matter above the vacuum,
$m_0=7.5$ MeV is the average current mass of
$u$ and $d$ quarks, $n^*$ is a constant related to the pion mass
$M_{\pi}=140$ MeV and the pion decay constant $F_{\pi}=93.2$ MeV by
\begin{equation} \label{nstar}
 n^* \equiv -\frac{2}{3} \langle\bar{q}q\rangle_0
  = \frac{M_\pi^2F_\pi^2}{3m_0} \approx 0.98\ \mbox{fm}^{-3}.
\end{equation}
Here we have used 
the GellMan-Oakes-Renner relation~\cite{GOR}
\begin{equation} \label{GOR}
-2 m_0\langle\bar{q}q\rangle_0 =M_{\pi}^2 F_{\pi}^2
\end{equation}
and the notations
$
\langle O\rangle_{n_{\text{b}}}
\equiv \langle n_{\text{b}}|O|n_{\text{b}}\rangle\
 \mbox{and}\
\langle O\rangle_0 \equiv \langle 0| O |0\rangle
$
for an arbitrary operator $O$.

From Eq.\ (\ref{qcCohen}), the usual procedure to calculate the
in-medium quark condensate will be to find an expression for
the energy density, carrying then the derivative with respect
to the current quark mass. However, it is usually difficult
to know the current quark mass dependence of model parameters,
e.g.\ meson masses, coupling constants, and cutoffs etc.,
without full understanding how effective models are related to
the underlying theory (QCD). To bypass this difficulty,
we suggest an equivalent mass method bellow.

%
%

 The QCD Hamiltonian density for the two-flavor case can be schematically
written as
\begin{equation}     \label{Hqcd}
  H_{\text{QCD}}=H_{\text{k}} + 2 m_{0}\bar{q}q + H_{\text{I}},
\end{equation}
where $H_{\text{k}}$ is the kinetic term, $H_{\text{I}}$ is the
interacting part,
and $m_0$ is the average current mass of $u$ and $d$ quarks.
Terms breaking flavor symmetry have been ignored.

The basic idea of the mass-density-dependent model of quark
matter is that the system energy can be expressed in the same form as
a proper non-interacting system. The strong interaction between quarks
is included within the appropriate variation of quark masses with density.
In order not to confuse with other mass concepts, we refer to such a
density-dependent mass as an equivalent mass.
Therefore, if we use the equivalent mass $m$, the system Hamiltonian
density should be replaced by an Hamiltonian density of the form
%
%
\begin{equation}    \label{Hequiv}
  H_{\text{eqv}}=H_{\text{k}} + 2m \bar{q}q,
\end{equation}
 where $m$ is the equivalent mass to be determined.
Obviously, we must require that the two Hamiltonian densities
$H_{\text{eqv}}$ and $H_{\text{QCD}}$ have the same expectation value
for any state $|\Psi\rangle$, i.e.,
\begin{equation}
  \langle{\Psi} | H_{\text{eqv}} |\Psi\rangle
 =\langle{\Psi} | H_{\text{QCD}} |\Psi\rangle.
\end{equation}
Applying this equality to the state $|n_{\text{b}}\rangle$\
with baryon number density $n_{\text{b}}$ and to the vacuum state $|0\rangle$,
respectively, taking then the difference, one has
\begin{equation} \label{Hrel}
  \langle H_{\text{eqv}}\rangle_{n_{\text{b}}}
 -\langle H_{\text{eqv}}\rangle_0
 =\langle H_{\text{QCD}}\rangle_{n_{\text{b}}}
 -\langle H_{\text{QCD}}\rangle_0.
\end{equation}

   Since we are considering a system with uniformly distributed
particles, or in other words, the density $n_{\text{b}}$ has nothing to do
with the space coordinates, we can write
$
\langle \Psi|m(n_{\text{b}})\bar{q}q|\Psi\rangle
=m(n_{\text{b}})\langle \Psi|\bar{q}q|\Psi\rangle.
$
This equality is especially obvious if we consider it in terms
of quantum mechanics: $|\Psi\rangle$ is a wave function with
arguments $n_{\text{b}}$ and coordinates, the expectation value is
nothing but an integration with respect to the coordinates.
Therefore, if $n_{\text{b}}$ does not depend on coordinates, the function
$m(n_{\text{b}})$ is also a coordinate-independent c-number, and can
naturally be taken out of the integration. However,
if $n_{\text{b}}$ is local, the case becomes
much more complicated and we will not consider it here.

Now we can solve Eq.\ (\ref{Hrel}) for $m$, and accordingly obtain
\begin{equation} \label{mdef}
m = m_0+\frac{\langle H_{\text{I}}\rangle_{n_{\text{b}}}
                   -\langle H_{\text{I}}\rangle_0}
{2(\langle\bar{q}q\rangle_{n_{\text{b}}}-\langle\bar{q}q\rangle_0)}
  \equiv m_0 + m_{\text{I}}.
\end{equation}

Therefore, considering our quarks as a free system, i.e.\
without interactions,  while keeping the
system energy unchanged, quarks should acquire an
equivalent mass of the form shown in Eq.\ (\ref{mdef}).
From this equation we see that the
equivalent mass $m$ includes two parts: one is the original mass
or current mass $m_{0}$, the other one is the interacting part
$m_{\text{I}}$.
Obviously the equivalent mass is a function of
both the quark current mass and the density. At finite temperature,
it depends also on the temperature as well. Here we consider
only zero temperature. Due to the quark confinement and asymptotic freedom,
one may naturally expect
\begin{equation} \label{mlim}
\lim_{n_{\text{b}}\rightarrow 0} m_{\text{I}}=\infty\
  \ \mbox{and}\ \
\lim_{n_{\text{b}}\rightarrow\infty} m_{\text{I}} =0.
\end{equation}

Because the Hamiltonian density $H_{\text{eqv}}$ has the same form
as that of a system of free particles with equivalent mass $m$,
the energy density of quark matter can be expressed as
\begin{equation}   \label{epsilon}
 \epsilon
= \frac{g}{2\pi^2} \int^{k_{\text{f}}}_0
   \sqrt{k^2+m^2}\ k^2 dk
=3mn_{\text{b}} F\left(\frac{k_{\text{f}}}{m}\right),
\label{epsi}
\end{equation}
where $g=2$(flavor)$\times$3(colors)$\times$2(spins) = 12 is the
degeneracy factor,
\begin{equation}
k_{\text{f}}=\left(\frac{18}{g}\pi^2n_{\text{b}}\right)^{1/3}
\end{equation}
is the Fermi momentum of quarks.

Substituting Eq.\ (\ref{epsilon}) into Eq.\ (\ref{qcCohen})
and carrying out the corresponding derivative, we have
\begin{equation}    \label{qc}
   \frac{\langle\bar{q}q\rangle_{n_{\text{b}}}}{\langle\bar{q}q\rangle_0}
 = 1 - \frac{n_{\text{b}}}{n^*} f\left(\frac{k_{\text{f}}}{m}\right)
       \frac{\partial m}{\partial m_0},
\end{equation}
where the chain relation
$$
\frac{\partial\epsilon}{\partial m_0}
=\frac{\partial\epsilon}{\partial m}
 \frac{\partial m}{\partial m_0}
$$
has been used. The functions $F$ in Eq.\ (\ref{epsilon})
and $f$ in Eq.\ (\ref{qc}) are, respectively, defined to be
\begin{eqnarray} \label{fdef}
F(x) &\equiv& \frac{3}{8}\left[
    x\sqrt{x^2+1}\left(2x^2+1\right)-\mbox{arcsinh}(x)
                         \right]/x^3,     \\
f(x) &\equiv& \frac{3}{2}\left[
        x\sqrt{x^2+1}-\ln\left(x+\sqrt{x^2+1}\right)
                         \right]/x^3
\end{eqnarray}
which have the following properties:
\begin{eqnarray}
& F(x)-x\frac{dF(x)}{dx}=f(x), &  \label{Ffrelat}  \\
&\lim\limits_{x\rightarrow 0} F(x)=1,  \ \
  \lim\limits_{x\rightarrow\infty} F(x)
   \rightarrow\frac{3}{4}\left(x+\frac{1}{x}\right), & \label{Flim}  \\
& \lim\limits_{x\rightarrow 0} f(x)=1,  \ \
 \lim\limits_{x\rightarrow\infty} f(x)\rightarrow\frac{3}{2x}. &
                             \label{flim}
\end{eqnarray}

Defining $\langle H_{\text{I}}\rangle$\ to be
the interacting energy density
$
\langle H_{\text{I}}\rangle_{n_{\text{b}}}-\langle H_{\text{I}}\rangle_0\
$
and considering Eq.\ (\ref{nstar}), the interacting part
of the equivalent mass,
 $m_{\text{I}}$ in Eq.\ (\ref{mdef}), can be re-written as
\begin{equation} \label{mI2}
m_{\text{I}}=\frac{\langle H_{\text{I}}\rangle/(3n^*)}
              {1-\langle\bar{q}q\rangle_{n_{\text{b}}}
                /\langle\bar{q}q\rangle_0}.
\end{equation}
Solving for the ratio
$\langle\bar{q}q\rangle_{n_{\text{b}}}/\langle\bar{q}q\rangle_0$,
this equation leads to
\begin{equation} \label{qc15}
\frac{\langle\bar{q}q\rangle_{n_{\text{b}}}}
     {\langle\bar{q}q\rangle_0}
=1-\frac{1}{3n^*}
   \frac{\langle H_{\text{I}}\rangle}{m_{\text{I}}}.
\end{equation}

 According to Eq.\
(\ref{Hqcd}), the total energy density of the quark matter system
can be expressed as
\begin{equation}
\epsilon
=\frac{g}{2\pi^2}\int_0^{k_{\text{f}}}
     \sqrt{k^2+m_0}k^2 dk
 +\langle H_{\text{I}}\rangle.
\end{equation}
The first term is the energy density without
interactions, the second term  $\langle H_{\text{I}}\rangle$
is the interacting part.
 On the other hand,
we have already expressed $\epsilon$ in Eq.\ (\ref{epsilon}).
So, replacing the $\epsilon$\ here with the right hand side of
Eq.\ (\ref{epsilon}), then dividing by $3n_{\text{b}}$,
we naturally get
\begin{equation} \label{EIexp}
\frac{\langle H_{\text{I}}\rangle}
    {3n_{\text{b}}}
= mF\left(\frac{k_{\text{f}}}{m}\right)
  -m_0F\left(\frac{k_{\text{f}}}{m_0}\right).
\end{equation}
Substituting this into Eq.\ (\ref{qc15})
gives
\begin{equation} \label{qc2}
\frac{\langle\bar{q}q\rangle_{n_{\text{b}}}}
     {\langle\bar{q}q\rangle_0}
=1-\frac{n_{\text{b}}}{n^*}
   \frac{1}{m_{\text{I}}}
   \left[
   mF\left(\frac{k_{\text{f}}}{m}\right)
  -m_0F\left(\frac{k_{\text{f}}}{m_0}\right)
   \right].
\end{equation}

Equating the right hand side of this equation
with that of Eq.\ (\ref{qc}), one can easily get
a fisrt order differential equation
\begin{equation}  \label{meq}
 \frac{\partial m}{\partial m_0}
=\frac{mF\left(k_{\mathrm{f}}/m\right)
       -m_{0}F\left(k_{\mathrm{f}}/m_0\right)}
      {(m-m_0) f\left(k_{\mathrm{f}}/m\right)}
\end{equation}
satisfied by the equivalent mass.
Eqs.\ (\ref{EIexp})-(\ref{meq}) seem similar to the Eqs.\ (23)-(25)
in Ref.~\cite{PengPLB548}. They are in fact different
by a factor of $N_{\text{f}}$, the number of quark flavors.
This difference will lead to different numerical results.

Now we have another formalism which can be used to
calculate the in-medium chiral condensate.
In principle, one can calculate the chiral condensate
at any finite density from Eqs.\ (\ref{qc2}) with the aid of
Eqs.\ (\ref{meq}) and (\ref{EIexp}), if one knows some information
on the interacting energy density $\langle H_{\mathrm{I}}\rangle$
from a realistic quark model. The obvious advantage of this
formalism is that one does not need to make assumptions on the
current mass derivatives of model parameters.
In the following, let's consider a simple model as an example.

If we denote the average distance between quarks by $\bar{r}$,
the interaction between quarks by $\mbox{v}(m_0,n_{\text{b}})$,
and assume each quark can only interact strongly with
another $N_0$ nearest quarks at any moment, because of
the saturation property of strong interactions, we can link
the interacting energy density $\langle H_{\text{I}}\rangle$
to density by
\begin{equation} \label{HI}
\langle H_{\text{I}}\rangle
=\frac{3}{2} N_0 n_{\text{b}} \mbox{v}(m_0,\bar{r}).
\end{equation}
This is obtained as follows. Suppose the total particle number
is $N$. The interacting energy due to particle $i$ is
$N_0\mbox{v}(m_0,n_{\text{b}})$, accordingly the total interacting
energy is
$
\frac{1}{2}\sum_{i=1}^{N} N_0\mbox{v}(m_0,n_{\text{b}})
=\frac{1}{2}N N_0\mbox{v}(m_0,n_{\text{b}})
$
(the extra factor 1/2 is for the correction of double counting).
Dividing this by the volume, we then have the interacting energy
density in Eq.\ (\ref{HI}) where the average inter-quark distance
$\bar{r}$ is linked to density through
\begin{equation}  \label{rbar}
\bar{r}=\frac{\xi}{n_{\text{b}}^{1/3}}.
\end{equation}

The average volume occupied by one particle is $1/(3n_{\text{b}})$.
If this volume is considered as a spherical ball of diameter $\bar{r}$,
one finds $\xi=(2/\pi)^{1/3}$. If, however, this volume is considered
as a cubic box of side $\bar{r}$, one has $\xi=1/3^{1/3}$.
In the former case, one divides the system into sub balls and place
a particle at the center of each small ball. In the latter case,
the system is divided into sub cubic boxes. Obviously there are
unoccupied spaces between balls in the former case, so we take the
latter value, i.e., $\xi=1/3^{1/3}$.

  Eq.\ (\ref{HI}) may not be an absolutely exact expression
for the interacting energy density. Some other components,
for example, the possible three-body interactions, have not been
included. However, they should be of higher orders in density,
and will not be considered here.
To compensate for these ignorance, to some extent,
 one may regard the $N_0$ as a free parameter. However, we will take
$N_0=2$ due to the fact that the quark has a trend to interact
strongly with other two quarks to form a baryon.
It will soon be shown that the relative chiral condensate
is independent of the concrete $N_0$ value at lower densities.

%

Substituting Eq.\ (\ref{HI}) into Eq.\ (\ref{EIexp}) gives
\begin{equation} \label{mvnorm}
\mbox{v}(m_0,\bar{r})
= \frac{2}{N_0} \left[
  mF\left(\frac{k_{\text{f}}}{m}\right)
  -m_0F\left(\frac{k_{\text{f}}}{m_0}\right)
                \right].
\end{equation}

Keeping in mind the second equalities in Eq.\ (\ref{mlim})
and in Eq.\ (\ref{Flim}), we have
\begin{equation} \label{vlim}
\lim\limits_{n_{\text{b}}\rightarrow\infty} \mbox{v}(m_0,\bar{r})=0.
\end{equation}
Note that we have said nothing about the form of the inter-quark
interaction v$(m_0,n_{\text{b}})$.

  At the same time, Eq.\ (\ref{qc15}) should be compatible with
Eq.\ (\ref{qc}). Comparing the right hand side of these
two equations, we get
\begin{equation} \label{mveq}
f\left(\frac{k_{\text{f}}}{m}\right) m_{\text{I}}
\frac{\partial m}{\partial m_0}
=\frac{N_0}{2}\mbox{v}(m_0,\bar{r}).
\end{equation}


Because Eq.\ (\ref{meq}) is a first order differential equation, we need
an initial condition at $m_0=m_0^*$ to get a definite solution.
 Let's suppose it to be
\begin{equation} \label{mini}
m(m_0^*,n_{\text{b}})=m(n_{\text{b}}).
\end{equation}
Usually, we will have
\begin{equation} \label{vini}
\mbox{v}(m_0,n_{\text{b}})|_{m_0=m_0^*}=v(\bar{r}),
\end{equation}
where $v(\bar{r})$ is the inter-quark interaction for
the special value $m_0^*$ of the quark current mass $m_0$.

Eq.\ (\ref{meq}) is difficult to solve analytically.
However, this can be done at lower densities.
For convenience, let's first prove that the equivalent mass
can be explicitly expressed through the inter-quark interaction as
\begin{equation} \label{mvexp}
m=m_0+\frac{\mbox{v}(m_0,\bar{r})
           }
           {\frac{2}{N_0} f\left(\frac{k_{\text{f}}}{m_0}\right)
            +\frac{\partial\mbox{v}(m_0,\bar{r})}{\partial m_0}
           }.
\end{equation}

In fact, taking the derivative with respect to $m_0$ at both
sides of Eq.\ (\ref{mvnorm}) gives
\begin{equation} \label{tem1}
f\left(\frac{k_{\text{f}}}{m}
 \right)\frac{\partial m}{\partial m_0}
=f\left(\frac{k_{\text{f}}}{m_0}\right)
 +\frac{N_0}{2}
  \frac{\partial\mbox{v}(m_0,\bar{r})}{\partial m_0}.
\end{equation}
Here Eq.\ (\ref{Ffrelat}) has been applied.
Then substituting Eq.\ (\ref{tem1}) into Eq.\ (\ref{mveq}), we
immediately get Eq.\ (\ref{mvexp}).

At lower densities, the Fermi momentum $k_{\text{f}}$ is small,
so the function $F(x)$ approaches to 1. Accordingly,
 Eq.\ (\ref{mvnorm}) leads to
\begin{equation}
m=m_0 
  +\frac{N_0}{2}\mbox{v}(m_0,\bar{r}).
\end{equation}
Replacing the left hand side of Eq.\ (\ref{mvexp})
with this expression, we get
\begin{equation}
\frac{N_0}{2}\frac{\partial\mbox{v}}{\partial m_0}
+f\left(\frac{k_{\text{f}}}{m_0}\right)=1.
\end{equation}
Integrating this equation under the initial condition
given in Eq.\ (\ref{vini}), we have
\begin{equation}
\mbox{v}(m_0,\bar{r})
=v(\bar{r}) + \int_{m_0^*}^{m_0}\frac{2}{N_0}
 \left[1-f\left(\frac{k_{\text{f}}}{m_0}\right)\right] dm_0.
\end{equation}
Then the interacting part of the equivalent mass is
\begin{equation} \label{mIlowexp}
m_{\text{I}}(m_0,n_{\text{b}})=\frac{N_0}{2} v(\bar{r})
+\int_{m_0^*}^{m_0}
 \left[1-f\left(\frac{k_{\text{f}}}{m_0}\right)\right] dm_0.
\end{equation}

 It is interesting to note that Eq.\ (\ref{mIlowexp}) can, because of
the dominant linear confining interaction, lead to
a quark mass scaling of the form
 $m_{\text{I}}\propto 1/n_{\text{b}}^{1/3}$ which is consistent
with the result in Ref.~\cite{PengPRC61}.

Substituting the above results for v and $m_{\text{I}}$
 into Eq.\ (\ref{qc15}) gives
\begin{equation} \label{qcL}
\frac{\langle\bar{q}q\rangle_{n_{\text{b}}}}{\langle\bar{q}q\rangle_0}
=1-\frac{n_{\text{b}}}{n^*}.
\end{equation}

This result is very similar to the famous model-independent
 result for the quark condensate in nuclear matter
\begin{equation} \label{qcnlin}
\frac{\langle\bar{q}q\rangle_{\rho}}{\langle\bar{q}q\rangle_0}
=1-\frac{\rho}{\rho^*}
 \ \mbox{with} \
\rho^*\equiv\frac{M_{\pi}^2F_{\pi}^2}{\sigma_{\text{N}}},
\end{equation}
which
was first proposed by Drukarev {\sl et al.}~\cite{Drukarev99},
and later re-justified by many
authors~\cite{Cohen91,Chanfray93,Lutz}.
We will also reproduce it in the next section
within the present approach.

Both Eqs. (\ref{qcL}) and (\ref{qcnlin}) mean a linear decreasing of
the condensate at lower densities. The ratio of their decreasing speeds
is
\begin{equation} \label{sigm0}
\frac{1/n^*}{1/\rho^*}=\frac{\rho^*}{n^*}
=\frac{3m_0}{\sigma_{\text{N}}}.
\end{equation}
If one takes  $m_0=(m_{u0}+m_{d0})/2=(5+10)/2=7.5$ MeV \cite{Gasser82}
and $\sigma_{\text{N}}=45$
MeV~\cite{Gasser91,Ericson87,Gensini80},
i.e., the sigma commutator is about six times the average current mass,
the decreasing speed in quark matter is merely half of that in nuclear matter.
If, however, one takes for $\sigma_{\text{N}}$ a value of about three
times $m_0$, as previously determined~\cite{Banerjee77},  the two decreasing
speeds are equal.  More recent investigations are in support of
a bigger value for $\sigma_{\text{N}}$~\cite{Sainio95,Gibbs98}.

In general, an explicit analytical solution for the condensate
 is not available, and we have to perform numerical calculations.
For a given inter-quark interaction $v(\bar{r})$,
we can first solve Eq.\ (\ref{mvnorm}) 
to obtain the initial condition
in Eq.\ (\ref{mini}) for the equivalent mass, then solve
the differential Eq.\ (\ref{meq}), and finally calculate the quark
condensate through Eq.\ (\ref{qc2}).

There are various expressions for $v(\bar{r})$ in literature,
e.g., the Cornell potential~\cite{Eichten75}, the Richardson
potential~\cite{Richardson79}, the so-called QCD potentials
\cite{Fischler77,Billoire80},
and purely phenomenological potentials~\cite{Quigg77,Martin80}, etc.
A common feature among them is that
they are all flavor-independent. This independence is
supported in a model-independent way by applying the inverse
scattering approach to extract a potential from the measured
spectra~\cite{Quigg81}.
Let's take a QCD-like interaction of the form
\begin{equation} \label{vqcdlike}
v(\bar{r})
=\sigma\bar{r}-\frac{4}{3}\frac{\alpha_s(\bar{r})}{\bar{r}}.
\end{equation}
The linear term $\sigma\bar{r}$ is the long-range confining part.
It is consistent with modern lattice simulations and string
investigations~\cite{lattice,string}.
%
%
 The second term
incorporates perturbative effects. To second order
in perturbation theory, one has~\cite{Fischler77,Billoire80}
\begin{equation}
\alpha_s(\bar{r})
=\frac{4\pi}{b_0\lambda(\bar{r})}
\left[
1-\frac{b_1}{b_0^2}\frac{\ln\lambda(\bar{r})}
{\lambda(\bar{r})}
 +\frac{b_2}{\lambda(\bar{r})}
\right]
\end{equation}
where~\cite{Igi86}
\begin{equation}
\lambda(\bar{r})
 \equiv \ln[(\bar{r}\Lambda_{\overline{ms}})^{-2}+b]
\end{equation}
with $b_0=(11N_{\text{c}}-2N_{\text{f}})/3$,
$b_1=[34N_{\text{c}}^2-N_{\text{f}}(13N_{\text{c}}^2-3)/N_{\text{c}}]/3$,
and $b_3=(31N_{\text{c}}-10N_{\text{f}})/(9b_0)$
for SU($N_{\text{c}}$) and $N_{\text{f}}$ flavors.

\begin{figure}
\centerline{\psfig{file=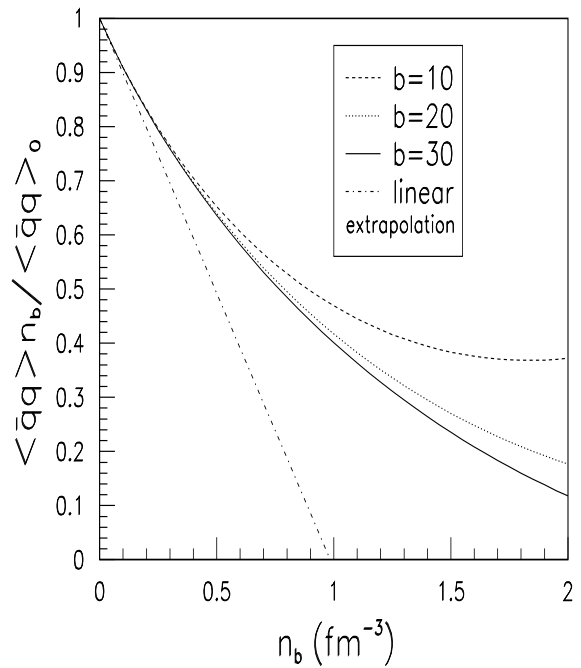,width=8cm,height=9cm}}
\vspace*{8pt}
\caption{
 Density dependence of the quark condensate in quark matter.
The straight line is the linear extrapolation of Eq.\ (\ref{qcL}).
The other three lines are for $m_0=7.5$ MeV, but for different
$b$ values, as indicated in the legend.
\label{qcnb}
              }
\end{figure}

\begin{figure}
\centerline{\psfig{file=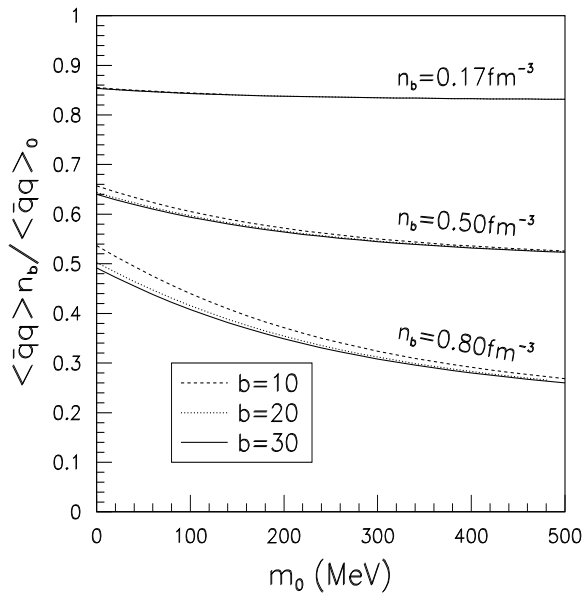,width=8cm,height=8cm}}
\vspace*{8pt}
\caption{
 Current mass dependence of the quark condensate
 in quark matter.
        }
\label{qcm0}
\end{figure}

Besides these constants, there are three parameters, i.e.,
$\sigma$, $\Lambda_{\overline{ms}}$, and $b$.
The QCD scale parameter is usually taken to be
$\Lambda_{\overline{ms}}=300$ MeV.
The value for the string tension $\sigma$ from potential
models varies in the range of 0.18---0.22 GeV$^2$~\cite{Veseli96},
and we here take $\sigma=0.2$ GeV$^2$.
As for the parameter $b$, we take three values, i.e., 10, 20, and 30,
in the reasonable range~\cite{Igi86}.
%
%
Because these parameters are determined from
heavy quark experimental data, the initial value $m_0^*$ is
taken to be 1500 MeV which is compatible with the mass of
heavy quarks. If we change this value to a smaller one,
the global features of the condensate will not be changed significantly.
Our numerical results are plotted in Fig.\
\ref{qcnb} and Fig.\ \ref{qcm0}.

Figure \ref{qcnb} shows the density behavior of the chiral
condensate in quark matter.
The straight line is the linear extrapolation of Eq.\ (\ref{qcL}).
It does not depend on the form of the inter-quark interaction
$v(\bar{r})$, and so, at this meaning, `model-independent'.
The other three lines are for $m_0=7.5$ MeV, but for different $b$
values, as indicated in the legend. At lower densities, the chiral
condensate decreases linearly with increasing densities. When the
density becomes higher, the decreasing speed is slowed.

Figure \ref{qcm0} shows the current mass dependence of the
chiral condensate in quark matter.
 From Eqs. (\ref{qc2}) and (\ref{meq}), it is not difficult to obtain
\begin{equation}
\frac{\partial}{\partial m_0}
      \left(
\frac{\langle\bar{q}q\rangle_{n_{\mathrm{b}}}}
     {\langle\bar{q}q\rangle_0}
       \right)
 =\frac{n_{\text{b}}}{{n^*}^2} F^*
   \left\{
   \frac{dn^*}{dm_0}
   -\frac{n^*}{m_{\text{I}}}
     \left[
    2-\frac{F^*}{f^*}-\frac{f_0^*}{F^*}
     \right]
   \right\}
\end{equation}
with
$
F^*(n_{\text{b}},m_{\text{I}})
\equiv \left[
        mF\left((k_{\text{f}}/m)\right)
        -m_0F\left((k_{\text{f}}/m_0)\right)
       \right]
       /{m_{\mathrm{I}}},
$
$f^*\equiv f(k_{\text{f}}/m)$,
and
$
f_0^*\equiv f(k_{\text{f}}/m_0).
$
The first term in the brace is due to the $m_0$ dependence of the vacuum
condensate while the second term is purely due to the density.
Here the numerical calculation only considers the second term to concentrate
on the density effect.
For the model Eq.\ (\ref{HI}) with Eq.\ (\ref{vqcdlike}),
the condensate drops when the current mass
becomes large. This effect is especially obvious
at higher densities. At lower densities, the condensate changes
little with increasing current mass. This is understandable:
because we have in fact assumed that the vacuum condensate
or the quantity $n^*$ in Eq.\ (\ref{nstar}) is constant,
the lower density condensate is nearly only a function of
the density.  If, however, the first term is negative,
i.e., the absolute value of the vacuum condensate decreases with
increasing $m_0$, as indicated in Ref.~\cite{Musakhanov02},
the in-medium condensate will drops more rapidly.
On the other hand, if the first term is positive enough,
the in-medium condensate will go up with increasing $m_0$.

Finally in this section, we would like to comment that
one can adopt any realistic model to calculate the
density dependence, and also current mass dependence if necessary,
  of the in-medium quark condensate
without making any further assumption by following the
formalism presented here, although we adopted a simple QCD-like
interaction as an example.  The procedure is as follows.
One first has the interacting energy density
$\langle H_{\text{I}}\rangle$\ from the selected realistic model
for a special current mass value,
gets the corresponding initial equivalent mass $m(n_{\text{b}})$ in
Eq.\ (\ref{mini}) from Eq.\ (\ref{EIexp}), 
obtains the equivalent mass $m(m_0,n_{\text{b}})$ by solving
Eq.\ (\ref{meq}), calculates then the quark condensate through
Eq.\ (\ref{qc2}). 

\section{The quark condensate in nuclear matter}
\label{qcnm}

In order to examine the correctness of the equivalent mass method,
we now apply it to the investigation of the quark condensate
in nuclear matter.

   To make a transit from quark matter to nuclear matter,
we suppose that nuclear matter can be well described by
some effective Hamiltonian density which can be schematically
expressed in the form
\begin{equation}   \label{hnm}
{\mathcal{H}}_{\text{NM}}
={\mathcal{H}}_{\text{k}}
 + 2 M_{\text{N}}\bar{N}N
 + {\mathcal{H}}_{\text{I}},
\end{equation}
where 
$M_{\text{N}}=938.926$ is the mass of nucleons in free space,
$\mathcal{H}_{\text{k}}$ is the kinetic term,
$\mathcal{H}_{\text{I}}$ is the interaction term between nucleons.
We also ignore the nucleon mass difference. In principle,
the following equality should hold for any nuclear state
$|\rho\rangle$\ with nucleon number density $\rho$:
\begin{equation} \label{hnhqcd}
   \langle\rho|{\mathcal{H}}_{\text{NM}}|\rho\rangle
  =\langle\rho|H_{\text{QCD}}|\rho\rangle.
\end{equation}
Here $H_{\text{QCD}}$ represents the QCD Hamiltonian density in
Eq.\ (\ref{Hqcd}). Proceeding in the similar way as in the
preceding section, we define an equivalent Hamiltonian density
\begin{equation}   \label{hst}
{\mathcal{H}}_{\text{eqv}}
 ={\mathcal{H}}_{\text{k}}
   + 2 M\bar{N}N,
\end{equation}
where $M$ is an equivalent mass to be determined
by the requirement that the energy of this Hamiltonian
is equal to the original interacting system, i.e.\
\begin{equation} \label{hnheqv}
   \langle\rho|\mathcal{H}_{\text{eqv}}|\rho\rangle
  =\langle\rho|\mathcal{H}_{\text{NM}}|\rho\rangle.
\end{equation}

Applying this equality to the nuclear state $|\rho\rangle$\
and the vacuum state $|0\rangle$, respectively,
and then taking the difference, we have
\begin{equation}
\langle\rho|{\mathcal{H}}_{\text{eqv}}|\rho\rangle
 -\langle 0|{\mathcal{H}}_{\text{eqv}}|0\rangle
=\langle\rho|{\mathcal{H}}_{\text{NM}}|\rho\rangle
 -\langle 0|{\mathcal{H}}_{\text{NM}}|0\rangle.
\end{equation}
Solving for $M$, this equation gives
\begin{eqnarray}
M &=& M_{\text{N}}
     +\frac{\langle\rho|{\mathcal{H}}_{\text{I}}|\rho\rangle
            -\langle{0}|{\mathcal{H}}_{\text{I}}|0\rangle}
           {2\left[\langle\rho|\bar{N}N|\rho\rangle
                   -\langle 0|\bar{N}N|0\rangle\right]}
                                        \label{mni}  \\
                &\equiv& M_{\text{N}} + M_{\text{I}}.
                                        \label{mmni}
\end{eqnarray}

   Obviously, $M_{\text{I}}$ comes from the interaction between
nucleons. It is generally a function of both the density $\rho$
and the quark current mass $m_0$.
 If the density $\rho$\ is not too high, $M_{\text{I}}$
is empirically much smaller than $M_{\text{N}}$.
For example, the binding energy per nucleon is less than 2\% of the
free nucleon mass at the nuclear saturation density. Because no interaction
exists when the density becomes zero, we naturally expect
\begin{equation}  \label{Mlim}
\lim_{\rho\rightarrow 0} M_{\text{I}}=0
 \ \ \mbox{or} \ \
\lim_{\rho\rightarrow 0} M=M_{\text{N}}.
\end{equation}

Using the equivalent mass, the energy density of the nuclear matter
can be expressed as
\begin{equation}    \label{epn}
  \epsilon_{\text{N}}=\frac{g_{\text{N}}}{2\pi^2}
    \int^{p_{\text{f}}}_0 \sqrt{p^2+M^2}\ p^2 dp,
\end{equation}
where $g_{\text{N}}=2$(isospins)$\times$ 2(spins) = 4
is the degeneracy factor,
$ p_{\text{f}}=\left(6\pi^2\rho/g_{\text{N}} \right)^{1/3}$
is the Fermi momentum of nucleons.

From the application of the Hellmann-Feynman theorem and the
GellMann-Oakes-Renner relation, as was done in Ref.~\cite{Cohen91},
one can easily get
\begin{equation}
  \frac{\langle\bar{q}q\rangle_{\rho}}{\langle\bar{q}q\rangle_0}
  = 1 - \frac{1}{3n^*}
        \frac{d\epsilon_{\text{N}}}{d{m_0}}.
                             \label{qcn1}
\end{equation}
Here $\epsilon_{\text{N}}$ is the energy density of nuclear matter,
and the quantity $n^*$ is given in Eq.\ (\ref{nstar}).
%
%

Substituting Eq.\ (\ref{epn}) into Eq.\ (\ref{qcn1}), we have
\begin{eqnarray}    \label{qcnucl}
   \frac{\langle\bar{q}q\rangle_\rho}{\langle\bar{q}q\rangle_0}
 &=& 1 - \frac{1}{3n^*}
        \frac{\partial\epsilon_{\text{N}}}{\partial M}
         \frac{\partial M}{\partial m_0}, \\
 &=& 1 - \frac{\rho}{3n^*}
        f\left(\frac{p_{\text{f}}}{M}\right)
        \frac{\partial}{\partial m_0}
        \left[
          M_{\text{N}} + M_{\text{I}}
        \right].
                            \label{qcn4}  \\
 &=&1-\frac{\rho}{\rho^*}
     f\left(\frac{p_{\text{f}}}{M}\right)
      \left[1+\frac{m_0}{\sigma_{\text{N}}}
    \frac{\partial M_{\text{I}}}{\partial m_0}
      \right].  \label{qcn5}
\end{eqnarray}
In obtaining the third equality, we have used the definition
of the pion-nucleon sigma term~\cite{Gasser81}
\begin{equation} \label{sigdef}
   \sigma_{\text{N}} = m_0 \frac{dM_{\text{N}}}{dm_0}
             \ \   \mbox{or}  \ \
    \frac{dM_{\text{N}}}{dm_0} = \frac{\sigma_{\text{N}}}{m_0}.
\end{equation}

\begin{figure}
\centerline{\psfig{file=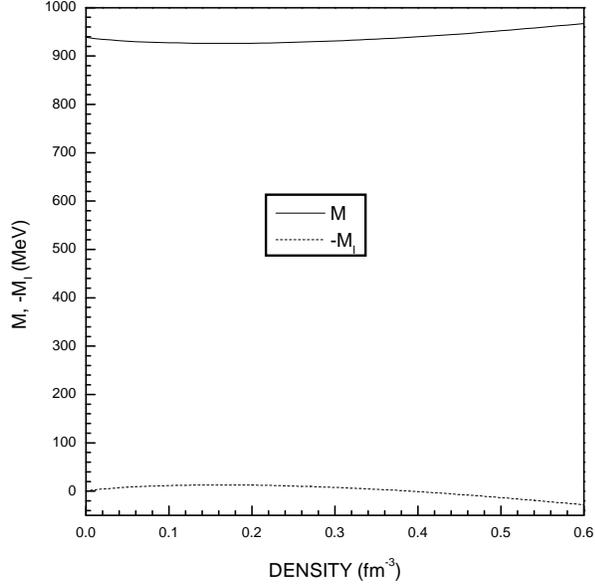,width=8cm}}
\vspace*{8pt}
\caption{
 Density dependence of the equivalent mass $M$ and its interaction
 part $M_{\text{I}}$ in the Dirac-Brueckner approach. It is
 obviously shown that $M_{\text{I}}$ is much smaller than $M_{\text{N}}$.
        }
\label{FigMIDB}
\end{figure}

The physical meaning of the equivalent mass is that nucleons should
have the equivalent mass if the system is free but with unchanged energy.
It is not difficult to prove that such an equivalent mass always exists.
In principle, if one obtain the energy density $\epsilon_{\text{N}}$
with some realistic models or even from QCD in the future,
the equivalent mass $M$ can be obtained by solving the equation
(\ref{epn}). For example, Fig.\ \ref{FigMIDB} shows the result
from the Dirac-Brueckner method with Bonn-B
potential~\cite{Machleidt}.

It should be emphasized that the equivalent mass $M$ is different from
the conventional effective mass $M^*$ of nucleons. For example,
at lower densities in Quantum Hadrondynamics (QHD)~\cite{Walecka},
the equivalent mass and effective mass can be expressed as
\begin{eqnarray}
M^*&=&M_{\text{N}}-\frac{g_\sigma^2}{M_\sigma^2}\rho, \label{mx1} \\
M &=& M_{\text{N}}-\frac{1}{2}\left(
  \frac{g_\sigma^2}{M_\sigma^2}-\frac{g_\omega^2}{M_\omega^2}
                             \right)\rho, \label{mx2}
\end{eqnarray}
where $M_\sigma=550$ MeV and $M_\omega=783$ MeV are, respectively,
the masses for the $\sigma$ and $\omega$ mesons,
and $g_{\sigma}=10.2$ and $g_\omega=12.55$ are the
corresponding coupling constants.

It is obvious from Eqs.\ (\ref{mx1}) and (\ref{mx2}) that
the conventional effective mass includes merely the contribution
from scalar mesons while the equivalent mass also includes the
contribution from other mesons.

Equations (\ref{mx1}) and (\ref{mx2}) are obtained as such.
The QHD Lagrangian is
\begin{eqnarray}
  L &=& \bar{\Psi}\left[\gamma^\mu (i\partial_\mu-g_\omega\omega_\mu)
    -(M_{N}-g_\sigma\sigma) \right] \Psi
     +\frac{1}{2} \partial_\mu \sigma \partial^\mu \sigma
                   \nonumber \\
   && -\frac{1}{2} M_\sigma^2\sigma^2
      -\frac{1}{4} F_{\mu\nu}F^{\mu\nu}
      +\frac{1}{2} M_\omega^2\omega_\mu\omega^\mu.     \label{LQHD}
\end{eqnarray}
At the mean-field level, we have the following energy
 density~\cite{Walecka}
\begin{eqnarray}
 \epsilon_{\text{QHD}}
 &=& g_{N} \int^{p_{\text{f}}}_0
         \sqrt{p^2+{M^*}^2}  \frac{d^3p}{(2\pi)^3}
                               \nonumber \\
&&+\frac{M_\sigma^2}{2g_{\sigma}^2}\left(M_{N}-M_{N}^*\right)^2
+\frac{g_\omega^2}{2M_\omega^2}\rho^2,       \label{EQHD}
\end{eqnarray}
where the effective mass $M^*$ is determined by the
self-consistent equation
\begin{equation}    \label{Meffeq}
  M^* = M_N-\frac{g_{\sigma}^2}{M_{\sigma}^2}
  \frac{g_N}{2\pi^2} \int_0^{p_{\text{f}}}
  \frac{M^*p^2dp}{\sqrt{p^2+{M^*}^2}}
\end{equation}
obtained by minimizing the energy density Eq.\ (\ref{EQHD}).

Taking the approximation $\sqrt{p^2+{M^*}^2}\approx M^*$ in
Eq.\ (\ref{Meffeq}) gives Eq.\ (\ref{mx1}) immediately.
As for Eq.\ (\ref{mx2}), it should be solved from
\begin{equation} \label{NQHD}
\epsilon_N(M,\rho)=\epsilon_{\text{QHD}}.
\end{equation}
Here the $\epsilon_N(M,\rho)$ is the right hand side (r.h.s.)
of Eq.\ (\ref{epn}) and $\epsilon_{\text{QHD}}$ is the r.h.s.\
of Eq.\ (\ref{EQHD}). Taking the approximation
\begin{equation} \label{rhocon}
 \sqrt{p^2+M^2} \approx M\
 \mbox{and}\
 \sqrt{p^2+{M^*}^2} \approx M^*
\end{equation}
in Eq.\ (\ref{NQHD}), Eq.\ (\ref{mx2}) is accordingly obtained.


\begin{figure}[hbt]
\centerline{\psfig{file=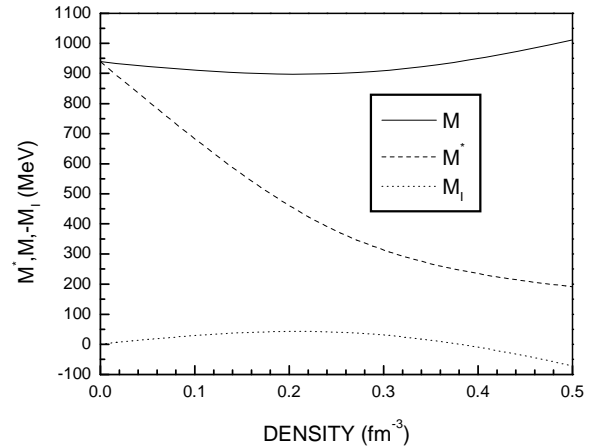,width=8cm}}
\vspace*{8pt}
\caption{The effective mass $M^*$,  the equivalent mass $M$,
      and the interaction mass $M_{\text{I}}$ vary with density in QHD.
         }
\label{FigQHD}
\end{figure}

Naturally, Eqs.\ (\ref{mx1}) and (\ref{mx2}) are valid only
to first order in density. For accurate relations, we have solved
Eqs.\ (\ref{Meffeq}) and (\ref{NQHD}) numerically.
The results are plotted in Fig.\ \ref{FigQHD}.

From Fig.\ \ref{FigMIDB} and Fig.\ \ref{FigQHD}, 
we see clearly that $M_{\text{I}}$ is much
smaller than $M_{\text{N}}$. Up to three times the normal nuclear
density, $M_{\text{I}}$ is less than 6 percent of $M_{\text{N}}$.
Also, when the density does not exceed the deconfinement density,
$M_{\text{I}}$ must be a continuous function of density. Therefore,
we ignore the second term in the bracket of Eq.\ (\ref{qcn4}), and get
\begin{equation}
\label{qcmore}
 \frac{\langle\bar{q}q\rangle_\rho}{\langle\bar{q}q\rangle_0}
  =1 - \frac{\rho}{\rho^*}
      f\left(\frac{p_{\text{f}}}{M}\right).
\end{equation}
Because the second term in the bracket of Eq.\ (\ref{qcn5})
has a factor $m_0/\sigma_{\text{N}}$ which is much smaller
than 1, we can also ignore the corresponding term and
get the same expression.

Due to the smallness of the ratio $p_{\text{f}}/M$ at lower
densities, the function $f(p_{\text{f}}/M)$ approaches to
unity. Accordingly, Eq.\ (\ref{qcmore}) becomes Eq.\
(\ref{qcnlin}), although Eq.\ (\ref{qcmore}) is in principle
more accurate. Naturally, if one wants to study the higher
density behavior of the quark condensate in nuclear matter,
the function $f(p_{\text{f}}/M)$ should be kept.
Also at higher densities, the omitted interacting part
$M_{\text{I}}$ should be taken into account.
In the following, we try to consider it in the mean-field level.

As has been found by Malheiro {\sl et al}.~\cite{Malheiro97},
the quark condensate can also be linked to the trace of
energy-momentum tensor of nuclear matter, i.e.\
\begin{equation} \label{qcmal}
 \frac{\langle\bar{q}q\rangle_\rho}{\langle\bar{q}q\rangle_0}
  =1 - \frac{1}{\rho^*}\frac{\epsilon_{\text{N}}-3P}{M_{\text{N}}}.
\end{equation}

To continue our discussion, we now adopt this equality
directly in our present model.
Because it is justified by Malheiro {\sl et al}.~\cite{Malheiro97}\
in the mean-field approximation of Walecka, ZM, and ZM3 models,
our treatment in the following are expected to be at least
valid at mean-field level.
In our case, the energy density $\epsilon_{\text{N}}$ is
related to the equivalent mass $M$ by Eq.\ (\ref{epn}) which,
after performing the integration, gives
\begin{equation} \label{epnexp}
\epsilon_{\text{N}}(M,\rho)
=M\rho F\left(\frac{p_{\text{f}}}{M}\right).
\end{equation}
Here the function $F$ is the same as in Eq.\ (\ref{fdef}).

The pressure $P$ can be obtained from Eq.\ (\ref{epnexp}), i.e.\
\begin{equation} \label{Pdef}
P=M\rho G\left(\frac{p_{\text{f}}}{M}\right)
  +\rho^2\frac{\partial M}{\partial\rho}
   f\left(\frac{p_{\text{f}}}{M}\right),
\end{equation}
where the function $f$ is given by Eq.\ (\ref{fdef}),
the function $G$ is defined to be
\begin{equation}
G(x)= \frac{1}{8} \left[
    x(2x^2-3)\sqrt{x^2+1}
           +3\mbox{arcsinh}(x)
                 \right]/x^3.  \label{defG}
\end{equation}

The second term on the r.h.s.\ of Eq.\ (\ref{Pdef})
is from the density dependence of the equivalent mass.
In obtaining the pressure $P$, one should be very careful.
Otherwise, inconsistencies will appear in this model.
For thermodynamic details, see our recent
publication~\cite{PengPRC62}.

Combining Eqs.\ (\ref{qcn4}) and (\ref{qcmal}) gives
\begin{equation} \label{Mpart}
\frac{\rho}{3n^*}f\left(\frac{p_{\text{f}}}{M}\right)
 \frac{\partial M}{\partial m_0}
=\frac{1}{\rho^*} \frac{\epsilon_{\text{N}}-3P}{M_{\text{N}}}.
\end{equation}
substituting Eqs. (\ref{epnexp}) and (\ref{Pdef})
into Eq.\ (\ref{Mpart}), we have
\begin{equation} \label{Mder1}
\frac{\rho^*}{3n^*}\frac{\partial M}{\partial m_0}
=\frac{M}{M_{\text{N}}}\left[
    \frac{F(p_{\text{f}}/M)-3G(p_{\text{f}}/M)}{f(p_{\text{f}}/M)}
           -\frac{3\rho}{M}\frac{\partial M}{\partial\rho}
                       \right].
\end{equation}

Because of the equalities,
\begin{equation}
\frac{F(x)-3G(x)}{f(x)}=1
      \ \ \mbox{and} \ \
\frac{\rho^*}{3n^*}=\frac{m_0}{\sigma_{\text{N}}},
\end{equation}
Eq.\ (\ref{Mder1}) can be cast to
\begin{equation} \label{Meq}
\frac{m_0}{\sigma_{\text{N}}}\frac{\partial M}{\partial m_0}
+\frac{3\rho}{M_{\text{N}}}\frac{\partial M}{\partial\rho}
=\frac{M}{M_{\text{N}}}.
\end{equation}

This partial differential equation is linear and of first order.
To find a special solution for our purpose, let's
expand the equivalent mass $M$ in a Taylor series at
zero density, i.e.\
\begin{equation} \label{Mexpand}
M(\rho,m_0)=\sum_i c_i\rho^i=c_0+c_1\rho+c_2\rho^2+c_3\rho^3.
\end{equation}
In principle, this expansion can have an arbitrarily large number of
higher order terms in density. But, for the moment, we only consider
terms up to the third order in density. We will discuss this problem
at the end of this section.

Substituting Eq.\ (\ref{Mexpand}) into Eq.\ (\ref{Meq}) leads to
\begin{equation} \label{coeff}
\frac{m_0}{\sigma_{\text{N}}}\sum_i {c_i}^{\prime}\rho^i
+\frac{3}{M_{\text{N}}}\sum_i ic_i\rho^i
=\frac{1}{M_{\text{N}}}\sum_i c_i\rho^i,
\end{equation}
where ${c_i}^{\prime}\equiv dc_i/dm_0$.
Comparing the coefficients in Eq.\ (\ref{coeff}), we have
\begin{equation} \label{cip}
{c_i}^{\prime}
=-\frac{\sigma_{\text{N}}}{m_0}
  \frac{3i-1}{M_{\text{N}}}c_i,
\end{equation}
or explicitly,
\begin{eqnarray}
{c_0}^{\prime}&=&\frac{\sigma_{\text{N}}}{m_0}
             \frac{c_0}{M_{\text{N}}},  \label{c0exp}  \\
{c_1}^{\prime}&=&-2\frac{\sigma_{\text{N}}}{m_0}
             \frac{c_1}{M_{\text{N}}},  \label{c1exp} \\
{c_2}^{\prime}&=&-5\frac{\sigma_{\text{N}}}{m_0}
             \frac{c_2}{M_{\text{N}}},  \label{c2exp} \\
{c_3}^{\prime}&=&-8\frac{\sigma_{\text{N}}}{m_0}
             \frac{c_3}{M_{\text{N}}}.  \label{c3exp}
\end{eqnarray}

From Eqs.\ (\ref{Mlim}) and (\ref{Mexpand}), we can easily
get $c_0=M_{\text{N}}$. In this case, Eq.\ (\ref{c0exp})
becomes ${c_0}^{\prime}=\sigma_{\text{N}}/m_0$, which is
consistent with Eq.\ (\ref{sigdef}).

It should be pointed out that the coefficients $c_i$
in the Taylor expansion (\ref{Mexpand}) are $m_0$-dependent.
In fact, we can integrate Eq.\ (\ref{cip}) and get
\begin{equation}
c_i=c_i(7.5)
 \left[\exp\left( {-\int_{7.5}^{m_0}
  \frac{\sigma_{\text{N}}}{M_{\text{N}}}
  \frac{dm_0}{m_0}}\right)
 \right]^{3i-1},
\end{equation}
where $c_i(7.5)$ is the value for $c_i$ at $m_0=7.5$ MeV which
is the value we take for the current mass of $u$ and $d$ quarks.

Now let's substitute Eq.\ (\ref{Mexpand}) into
Eq.\ (\ref{qcn4}) or (\ref{qcn5}) and get
\begin{equation}
 \frac{\langle\bar{q}q\rangle_\rho}{\langle\bar{q}q\rangle_0}
=1-\frac{\rho}{\rho^*} f\left(\frac{p_{\text{f}}}{M}\right)
   \left[1+\frac{m_0}{\sigma_{\text{N}}}
     \sum_{i=1}^3 {c_i}^{\prime}\rho^i \right].
\end{equation}
On applying Eq.\ (\ref{cip}), we finally have
\begin{equation} \label{qcnmfin}
 \frac{\langle\bar{q}q\rangle_\rho}{\langle\bar{q}q\rangle_0}
=1-\frac{\rho}{\rho^*} f\left(\frac{p_{\text{f}}}{M}\right)
   \left[1-\frac{1}{M_{\text{N}}}
            \sum_{i=1}^3 (3i-1)c_i\rho^i\right].
\end{equation}

For figure-plotting convenience, let's change the coefficients
$c_i$ (i=1,2,3) to $a_i$ by defining
\begin{equation}  \label{ca}
a_i=\frac{{\rho_0}^i}{M_{\text{N}}}c_i
  \ \mbox{or} \
c_i=\frac{M_{\text{N}}}{{\rho_0}^i} a_i.
\end{equation}
Accordingly, Eqs.\ (\ref{Mexpand}) and (\ref{qcnmfin})
become, respectively,
\begin{eqnarray}
\frac{M}{M_{\text{N}}}
  &=& 1+a_1 \frac{\rho}{\rho_0}
       +a_2 \left(\frac{\rho}{\rho_0}\right)^2
       +a_3 \left(\frac{\rho}{\rho_0}\right)^3, \label{Mexpa} \\
\frac{\langle\bar{q}q\rangle_\rho}{\langle\bar{q}q\rangle_0}
  &=& 1-\frac{\rho}{\rho^*} f\left(\frac{p_{\text{f}}}{M}\right)
        \left[1- 
 \sum_{i=1}^3 (3i-1)a_i\left(\frac{\rho}{\rho_0}\right)^i\right].
                                 \label{qccalculate3}
\end{eqnarray}
To obtain the values of the coefficients $a_1, a_2$, and $a_3$,
we can use the following known experimental properties
 of nuclear matter
\begin{eqnarray}
&\left.\frac{\epsilon_{\text{N}}(M,\rho)}{\rho}\right|
   _{\rho=\rho_0}=M_{\text{N}}+E_{\text{b}},& \label{nucprop1}\\
&\frac{d}{d\rho}\left.\left[\frac{\epsilon_{\text{N}}(M,\rho)}{\rho}
            \right]\right|_{\rho=\rho_0}=0, & \label{nucprop2}\\
&9\rho_0^2\frac{d^2}{d\rho^2}\left.\left[
                     \frac{\epsilon_{\text{N}}(M,\rho)}{\rho}
                                   \right]
                  \right|_{\rho=\rho_0}=K.&  \label{nucprop3}
\end{eqnarray}
Here
$\rho_0=0.17$ fm$^{-3}$ is the normal nuclear saturation density,
$E_{\text{b}}=-16$ MeV is the nuclear binding energy.
As for the incompressibility $K$, it is not so definite, maybe
ranging from 150 to 350 MeV. The expression
$\epsilon_{\text{N}}(M,\rho)/\rho=M F(p_{\text{f}}/M)$ is the
energy per nucleon.
For a given $K$, we can solve Eqs.\ (\ref{nucprop1}-\ref{nucprop3})
and get the values for $M^{(i)}$ ($i$=0,1,2):
\begin{eqnarray}
M^{(0)} &\equiv& \left.M\right|_{\rho=\rho_0},  \label{dM01} \\
M^{(1)} &\equiv& \left.\frac{\partial M}{\partial\rho}
            \right|_{\rho=\rho_0},         \label{dM02} \\
M^{(2)} &\equiv& \left.\frac{\partial^2M}{\partial\rho^2}
            \right|_{\rho=\rho_0}.          \label{dM03}
\end{eqnarray}
Substituting the expansion (\ref{Mexpa}) into the right hand side
of Eqs.\ (\ref{dM01}-\ref{dM03}) and then solving for
 $a_1, a_2$, and $a_3$, we have
\begin{eqnarray}
a_1
 &=&\frac{-1}{M_{\text{N}}}
     \left[3(M_{\text{N}}-M^{(0)})
          +2M^{(1)}\rho_0
          -\frac{1}{2}M^{(2)}{\rho_0}^2
    \right], \\
a_2
 &=&\frac{1}{M_{\text{N}}}
     \left[3(M_{\text{N}}-M^{(0)})
          +3M^{(1)}\rho_0
          -M^{(2)}{\rho_0}^2
    \right], \\
a_3
 &=&\frac{-1}{M_{\text{N}}}
     \left[M_{\text{N}}-M^{(0)}
          +M^{(1)}\rho_0
          -\frac{1}{2}M^{(2)}{\rho_0}^2
    \right].
\end{eqnarray}
The quark condensate can then be calculated
 by Eq.\ (\ref{qccalculate3}).

\begin{figure}
\centerline{\psfig{file=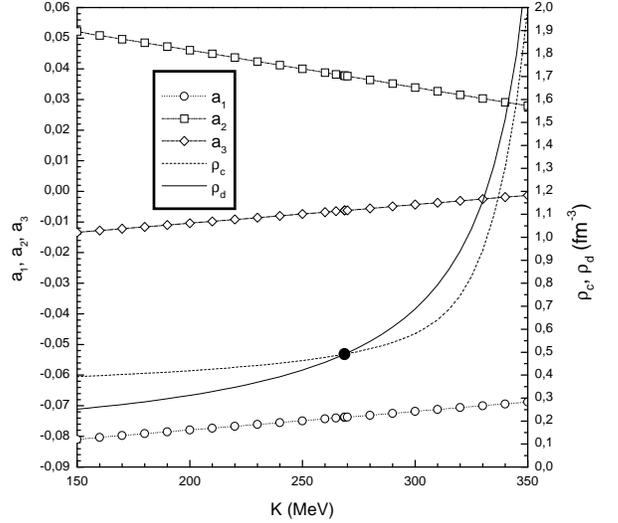,width=8cm}}
\vspace*{8pt}
\caption{The coefficients $a_1$, $a_2$, and $a_3$ vary
 with the incompressibility $K$. The dotted line gives
the corresponding critical density $\rho_{\text{c}}$ at
which the quark condensate becomes zero. The full line
is the deconfinement density $\rho_{\text{d}}$ after which
 the deconfinement phase transition is expected to occur. }
\label{figak}
\end{figure}

In Fig.\ \ref{figak}, we give, on the left axe,
the coefficients $a_1, a_2$, and $a_3$ as a function of $K$. Another
two lines on the right axe are to be explained a little later.

\begin{figure}
\centerline{\psfig{file=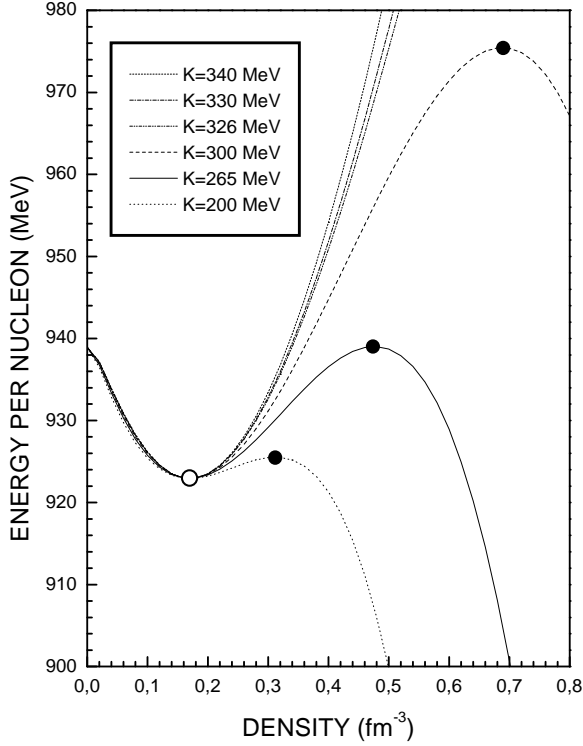,width=8cm}}
\vspace*{8pt}
\caption{Equation of state of nuclear matter for different
incompressibilities in the present approach. The open circle
corresponds to the saturation. After the solid circle, nuclear matter
will automatically contract, and the deconfinement phase is
expected to occur.}
\label{figeos}
\end{figure}

Fig.\ \ref{figeos} shows the equation of state of nuclear matter
for typical incompressibilities. The open circle corresponds to
the saturation which is a mechanically stable point,
so every line passes it by. The points marked with a solid circle
are also in mechanical equilibrium, but they are not stable.
 When the density is
between the open and solid circle, the pressure is positive.
However, when the density exceeds the full circle, the pressure
is drastically negative, so the system will automatically contract
and the deconfinement phase is expected to appear. We therefore
use $\rho_{\text{d}}$ to remember this density and plot it
in Fig.\ \ref{figak} with a solid line on the right axe.
We can also see that the incompressibility $K$ should be
 no less than 265 MeV because nuclear matter will be less
bounded than 16 MeV on the right side when $K<265$, i.e.\
the difference of the energy corresponding to the solid and open
circles is smaller than 16 MeV if $K$ is less than 265 MeV.

\begin{figure}
\centerline{\psfig{file=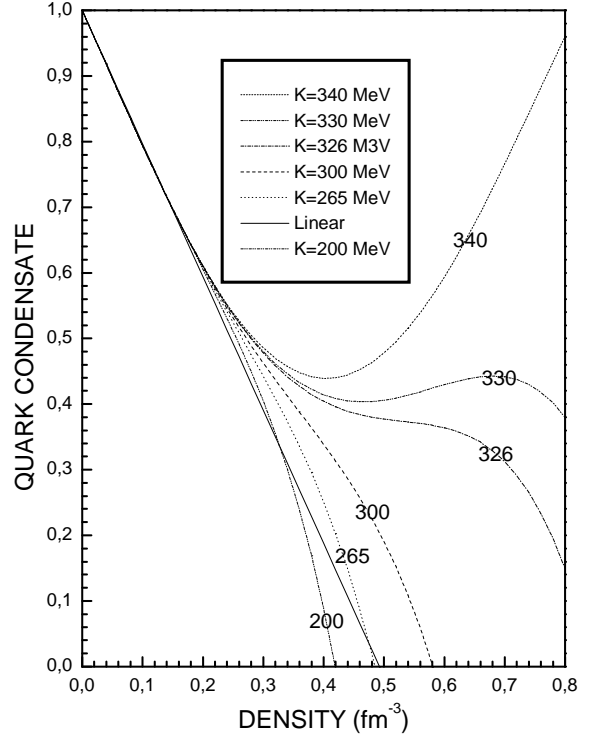,width=8cm}}
\vspace*{8pt}
\caption{
 The Quark condensate in nuclear matter to its vacuum value
 in the present model.
 When the incompressibility $K$ is in the range 265--326 MeV,
 the condensate decreases monotonously with increasing density.
 When $K$ is bigger than 326 MeV, there appears a hill where
 the chiral symmetry breaking might be enhanced. In both case,
 the critical density for the condensate becomes zero is higher
 than the linear extrapolation.
        }
\label{figqcnm}
\end{figure}

The quark condensate is given in Fig.\ \ref{figqcnm} for
typical values of $K$. For $K>265$, the critical density
$\rho_{\text{c}}$ where the quark condensate becomes zero
is generally bigger than that obtained by extrapolating
the model-independent result of Eq.\ (\ref{qcnlin}) (indicated
 with the word 'Linear' in the figure legend).
The variation of the critical density
with the incompressibility is shown in Fig.\ \ref{figak} with
a dotted line on the right axe. A noticeable feature is that,
for $K>326$, there appears a hill where the chiral symmetry breaking
might be enhanced.

As mentioned before, the Taylor expansion Eq.\ (\ref{Mexpand}) can
have infinitely large number of terms in density. However,
it has been obviously shown in Fig.\ \ref{figak} that
\begin{equation}
|a_1|>|a_2|>|a_3|.
\end{equation}
This is an indication that the higher the term's order is,
the less important its contribution. This is why we have
only taken to the third order in the expansion Eq.\ (\ref{Mexpand}).
However, it is very easy to extend to even higher orders.
For this purpose, we should first express the equivalent mass as
\begin{equation}
M(\rho,m_0)=\sum_{i=0}^J c_i\rho^i,
\end{equation}
where $J$ is an integer. It's value depends on to which order
we would like to expand. In the case we have just treated, $J=3$.

The initial values for
\begin{equation}
M^{(i)}\equiv \left.\frac{\partial^i M}{\partial\rho^i}
            \right|_{\rho=\rho_0} \ \ \ \  (i=0,\cdots,J-1)
\end{equation}
are solved from
\begin{equation}
(3\rho_0)^i \frac{d^i}{d\rho^i}
   \left.\left[\frac{\epsilon_{\text{N}}(M,\rho)}{\rho}
   \right]\right|_{\rho=rho_0}
=K_i,  \ \ \ (i=0 \cdots J-1),
\end{equation}
where $K_0=M_{\text{N}}+E_{\text{b}}, K_1=0, K_2=K$. If $J>3$,
we should also know the information on $K_3, K_4$ {\sl et al.}.
The coefficients $c_i$ $(i=1 \dots J)$ are obtained by solving
\begin{equation}
\sum_{i=k}^J \frac{i!}{(i-k)!}c_i{\rho_0}^i
=M^{(k)},  \ \ \ (k=0 \cdots J-1).
\end{equation}
And the quark condensate is
\begin{equation}
\frac{\langle\bar{q}q\rangle_\rho}{\langle\bar{q}q\rangle_0}
 = 1-\frac{\rho}{\rho^*} f\left(\frac{p_{\text{f}}}{M}\right)
     \left[1- 
 \sum_{i=1}^J (3i-1)a_i\left(\frac{\rho}{\rho^0}\right)^i\right].
\end{equation}
Here the relation between $a_i$ and $c_i$ is still given by
Eq.\ (\ref{ca}).

\section{Summary}
\label{dis}

  We have presented a new formalism for the calculation of
the in-medium chiral condensates. The key point of this scheme
is to obtain an equivalent mass by solving a differential
equation. The main advantage is that one does not need to make
further assumptions on the derivatives of model parameters
with respect to the quark current mass.

 As an application of this method,
we have successfully reproduced the normally accepted
model-independent result in nuclear matter. We also
showed that there is a similar expression for the chiral
condensate in quark matter.
It is found that the pion-nucleon sigma term is
six times the average light quark current mass
if the two decreasing speed is equal.

  For quark matter at higher densities,  we adopted a QCD-like
interaction to solve our differential equation, which
shows that the condensate decreases linearly at lower densities.
However, the decreasing speed is slowed at higher densities, compared
with the linear extrapolation.
For some values of the parameters, it can finally vanish,
while for other values it may saturate and even increase.

  We have also shown that the incompressibility $K$ should not be
less than 265 MeV to give the correct nuclear binding energy in
this model.
When $K$ is in the range of 265 to 326 MeV, the quark condensate
monotonously decreases with increasing density.
However, the critical density where the
condensate vanishes is, generally, higher than that obtained
by linear extrapolation. When $K$ is greater than 326 MeV,
a hill for the condensate as function of density will appear,
enhancing the chiral symmetry breaking, vanishing then for
even higher densities.

\section*{Acknowledgments}

The authors would like to thank Prof.\ R.\ Machleidt for providing
his latest version of the Dirac-Brueckner code.
The supports from the National Natural Science Foundation of China
 (19905011, 10135030, 10275037, 10075057, 10147208),
the CAS president foundation  (E-26), the CAS key project (KJCX2-SW-N02),
the university doctoral program foundation of the ministry of education
of China (20010055012),
and the FONDECYT of Chile (Proyecto 3010059 and 1010976)
are also gratefully acknowledged.

\end{document}